\documentclass[%
 reprint,
superscriptaddress,
showpacs,preprintnumbers,
 amsmath,amssymb,
 aps,
 prl,
]{revtex4-1}

\draft 
\usepackage{graphicx}
\begin{document}


\title{Resonantly excited betatron hard X-Rays from Ionization Injected Electron Beam in a Laser Plasma Accelerator} 


\affiliation{Beijing National Laboratory of Condensed Matter Physics, Institute of Physics, CAS, Beijing 100190, China}
\affiliation{Institute of High Energy Physics, CAS, Beijing 100049, China}
\affiliation{Key Laboratory for Laser Plasmas (MOE) and Department of Physics and Astronomy, Shanghai Jiao Tong University, Shanghai 200240, China}
\affiliation{IFSA Collaborative Innovation Center, Shanghai Jiao Tong University, Shanghai 200240, China}
\affiliation{SUPA, Department of Physics, University of Strathclyde, Glasgow G4 0NG, United Kingdom}

\author{K. Huang}
\affiliation{Beijing National Laboratory of Condensed Matter Physics, Institute of Physics, CAS, Beijing 100190, China}
\author{L. M. Chen}
\email{lmchen@iphy.ac.cn}
\affiliation{Beijing National Laboratory of Condensed Matter Physics, Institute of Physics, CAS, Beijing 100190, China}
\affiliation{IFSA Collaborative Innovation Center, Shanghai Jiao Tong University, Shanghai 200240, China}
\author{Y. F. Li}
\affiliation{Beijing National Laboratory of Condensed Matter Physics, Institute of Physics, CAS, Beijing 100190, China}
\author{D. Z. Li}
\affiliation{Institute of High Energy Physics, CAS, Beijing 100049, China}
\author{M. Z. Tao}
\affiliation{Beijing National Laboratory of Condensed Matter Physics, Institute of Physics, CAS, Beijing 100190, China}
\author{M. Mirzaie}
\affiliation{Key Laboratory for Laser Plasmas (MOE) and Department of Physics and Astronomy, Shanghai Jiao Tong University, Shanghai 200240, China}
\author{Y. Ma}
\affiliation{Beijing National Laboratory of Condensed Matter Physics, Institute of Physics, CAS, Beijing 100190, China}
\author{J. R. Zhao}
\affiliation{Beijing National Laboratory of Condensed Matter Physics, Institute of Physics, CAS, Beijing 100190, China}
\author{M. H. Li}
\affiliation{Beijing National Laboratory of Condensed Matter Physics, Institute of Physics, CAS, Beijing 100190, China}
\author{M. Chen}
\affiliation{Key Laboratory for Laser Plasmas (MOE) and Department of Physics and Astronomy, Shanghai Jiao Tong University, Shanghai 200240, China}
\author{N. Hafz}
\affiliation{Key Laboratory for Laser Plasmas (MOE) and Department of Physics and Astronomy, Shanghai Jiao Tong University, Shanghai 200240, China}
\author{T. Sokollik}
\affiliation{Key Laboratory for Laser Plasmas (MOE) and Department of Physics and Astronomy, Shanghai Jiao Tong University, Shanghai 200240, China}
\author{Z. M. Sheng}
\affiliation{Key Laboratory for Laser Plasmas (MOE) and Department of Physics and Astronomy, Shanghai Jiao Tong University, Shanghai 200240, China}
\affiliation{IFSA Collaborative Innovation Center, Shanghai Jiao Tong University, Shanghai 200240, China}
\affiliation{SUPA, Department of Physics, University of Strathclyde, Glasgow G4 0NG, United Kingdom}
\author{J. Zhang}
\email{jzhang1@sjtu.edu.cn}
\affiliation{Key Laboratory for Laser Plasmas (MOE) and Department of Physics and Astronomy, Shanghai Jiao Tong University, Shanghai 200240, China}
\affiliation{IFSA Collaborative Innovation Center, Shanghai Jiao Tong University, Shanghai 200240, China}



\date{\today}

\begin{abstract}
A new scheme for bright hard x-ray emission from laser wakefield electron accelerator is reported, where pure nitrogen gas is adopted. Intense Betatron x-ray beams are generated from ionization injected K-shell electrons of nitrogen into the accelerating wave bucket. The x-ray radiation shows synchrotron-like spectrum with total photon yield 8$\times$10$^8$/shot and $10^8$ over 110keV. In particular, the betatron hard x-ray photon yield is 10 times higher compared to the case of helium gas under the same laser parameters. Particle-in-cell simulation suggests that the enhancement of the x-ray yield results from ionization injection, which enables the electrons to be quickly accelerated to the driving laser region for subsequent betatron resonance. Employing the present scheme,the single stage nitrogen gas target could be used to generate stable high brightness betatron hard x-ray beams.
\end{abstract}


\maketitle 

Ultrafast x-ray sources have tremendous applications in time resolved x-ray diffraction and x-ray absorption spectroscopy to study the transient properties of condensed matter and biological structures\cite{eric1,eric2}, which have been realized so far mainly by use of X-ray free electron lasers (XFELs) with high brightness\cite{eric3,eric4}. However, XFELs are huge facilities accessible to limited users. On the other hand, with the development of femtosecond high power lasers \cite{eric5}, laser plasma x-ray sources are becoming increasingly attractive due to their compactness and natural synchronization of drive lasers and produced x-ray sources. X-ray emission from laser plasma interactions, such as K¦Á x-ray emission, nonlinear Thomson scattering and betatron x-ray sources, have been intensively studied \cite{eric6,eric7,eric8,eric9,eric10,eric11,eric12,eric13,eric14}. In particular, the betatron x-ray emission from electron oscillations in the laser wakefield acceleration (LWFA) is a promising source for its high spatial coherence\cite{eric12}, sound photon yield ($>10^{8}$/shot) and high photon energy (up to MeV)\cite{eric13}.

It is well-known that in three-dimensional geometry, trapped electrons inside the laser wakefield will not only be accelerated longitudinally\cite{eric15}, but also will undergo betatron oscillations due to the presence of transverse electric and magnetic fields in the wakefield\cite{eric14,eric16}. The oscillation period of the betatron motion in LWFA is thousand times smaller than that of the conventional magnetic wiggler. Therefore, betatron x-ray generated from LWFA can reach hard x-ray or $\gamma$-ray region\cite{eric17}. The betatron radiation is emitted in regimes that are distinguished by the wiggling strength parameter,which is defined as $a_\beta   = 2\pi \gamma r_\beta  /\lambda _\beta  $, where $r_\beta$ and $\lambda_\beta$ are the betatron oscillation amplitude and wavelength, respectively\cite{eric18}. In the case of LWFA where $a_{\beta}\gg1$ , the emitted x-ray beam has a synchrotron like broadband spectrum which can be expressed as$S(E,E_c )\sim N_\beta  \frac{{3e^2 }}{{2\pi ^3 \hbar c\varepsilon _0 }}\gamma ^2 (E/E_c )^2  \cdot K_{2/3}^2 (E/E_c )$, where$N_\beta$ is the number of oscillations, $K_{2/3}$ is a modified Bessel function of the second kind and $E_c  = 3\hbar a_\beta  \gamma ^2 \omega _\beta$ is the critical energy ($\hbar$ is the reduced Planck constant and $\omega_\beta$ is the betatron frequency). The spectrum peaks at E$\sim$0.5$E_c$, beyond which the radiation tends to decay exponentially\cite{eric12}.
 \begin{figure}[!htb]
\includegraphics[width=1\linewidth]{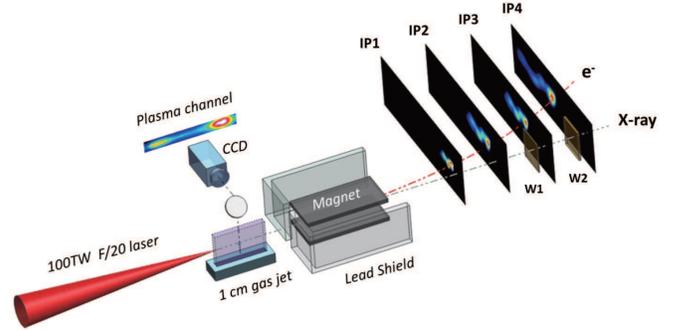}
\caption{\label{1} Experimental setup. All IPs are wrapped with 13$\mu$m Al foil. Two tungsten filter with thickness of W1=25$\mu$m and W2=56$\mu$m are put respectively between IP2 and IP3, IP3 and IP4 on the laser axis to attenuate x-ray beam. The consecutive IPs show the dispersed electron bunch. A sample of the plasma channel is shown in the inset above CCD.}
\end{figure}

The study of electron acceleration and resultant betatron x-ray emission from LWFA usually utilize low Z gases working in the self-injection regime. The betatron photon properties vary with different laser parameters and plasma densities in experiments\cite{eric11,eric12,eric13,eric14}. In different conditions, the betatron source size can be reached to a few $\mu$m level\cite{eric12,eric19} and the photon energy can get into gamma-ray region\cite{eric13} independently. However, for real applications, one needs production of x-ray emission including enough photon numbers and suitable radiation spectra. Therefore, controlling of the accelerated electron energy, total charge as well as their betatron oscillation dynamics is highly important. Ionization injection in the LWFA has the advantage of reduced laser intensity thresholds and better injection control as compared to self-injection, even though the energy spread of the produced beams still need to be continuously improved\cite{eric20,eric21,eric22,eric23,eric24}. Until now, betatron x-ray emission via ionization injected electron beams has not been investigated.

In this Letter, we report the first study of bright hard x-rays based upon ionization-injected electron beams accelerated in LWFA via betatron oscillations. Highly collimated hard x-rays with a photon flux of $8\times10^8$ per shot and with $10^8$ photons over 110keV have been produced with a pure nitrogen gas jet irradiated with 100TW laser pulses. This yield is about 10 times higher than that obtained with helium gas under similar laser conditions, and much higher than other experiment results reported working in the self-injection mode. Particle-in-cell (PIC) simulations suggest that the enhanced betatron photon energy and photon flux are due to ionized injection and betatron resonant oscillation in the laser fields.

The experiment was carried out using the hundred TW laser system at the Key Laboratory for Laser Plasmas in Shanghai Jiao Tong University. In the experiment, the system delivered 40fs pulses with energy up to 3J. The pulses were focused by a f/20 off-axis-parabola onto a 1.2mm$\times$10mm supersonic gas jet. The focal spot has a $1/e^{2}$ radius of $w_0$=21$\mu$m containing 50\% energy. The resultant laser peak intensity was up to 1.0$\times$10$^{19}$W/cm$^2$, corresponding to normalized vector potential $a_{0}=2.2$. A top-view system was set to monitor Thomson-scattering. The electron beams emitted from the gas jet were dispersed by a 16cm-long dipole magnet with magnetic field strength 0.98T. A combo of 4 image plates (IP) (Fuji Film SR series) covered with 12$\mu$m Al foil was set behind the magnet to record the electron and x-ray signal simultaneously. The IP had been calibrated to measure the electron beam charge and betatron x-ray photon yield\cite{eric25,eric26}. The electron energy was determined by different displacements on the consecutive IPs as in Ref.\cite{eric21}. Since IP (SR series) contains a 160¦Ìm thick ion layer, it was also be used as x-ray filters as well as tungsten filters, as shown in Figure 1. Thus,the photo-stimulated luminescence (PSL) values deposited on each IP correspond to x-ray photon energy above 7 keV, 36keV, 52 keV and 110 keV, respectively. For comparison, nitrogen and helium gases had been used in the same experiment. Since there is a large ionization potential gap between L-shell electrons and K-shell electrons of the nitrogen atom, the charge state for the background nitrogen ions is assumed to be 5. The measurement of gas density generated from this nozzle has been performed previously\cite{eric27}.

\begin{figure}
\includegraphics[width=0.9\linewidth]{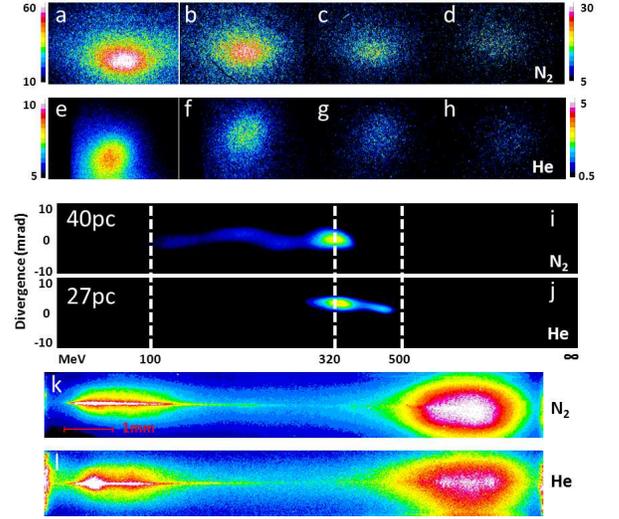}
\caption{\label{2} Experimental results. Frames (a-d) and (e-h) are the betatron x-ray signals with nitrogen and helium gases, respectively, on IP1-4; Frames (i) and (j) show the electron energy spectra with nitrogen and helium gases, which are given in the same color scale; Frames (k) and (l) show the plasma channel images with nitrogen and helium gases, respectively.}
\end{figure}

\begin{figure}[!htb]
\includegraphics[width=1\linewidth]{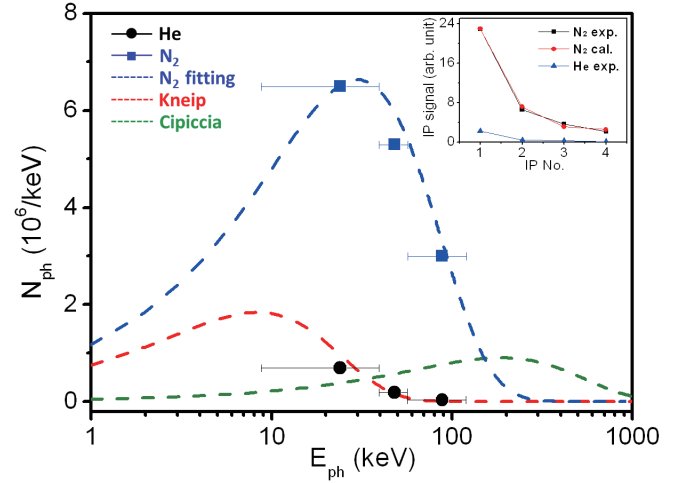}
\caption{\label{3} X-ray spectrum analysis. The blue squares and black circles denote the measured average photon number produced with nitrogen and helium gases, respectively. The blue dashed line shows the fitted spectrum for nitrogen gas with the critical energy of 75keV. Red and green dashed lines represent the x-ray spectra by Kneip et al. [12] and Cipiccia et al. [13]; Inset in the figure shows the x-ray signal strength on consecutive IPs. The black squares and red circles represent the experimental signal and the calculated value using synchrotron radiation distribution, respectively. The blue triangle shows the experimental x-ray signal with helium gas.}
\end{figure}
Figure 2 shows the experimental betatron x-ray profile(a-d,e-h), corresponding electron spectra signal(i,j) and laser plasma channel length(k,l) for different gas species. By integration of the x-ray signal, we find that the betatron x-ray emission from nitrogen plasma(a-d) is 10-fold than that from the helium plasma(e-h) for x-ray $>$ 7keV. The profiles of the betatron signals from the nitrogen gas exhibit a clear elliptical structure with the divergence angles of 6mrad and 3mrad in FWHM along the long axis and short axis, respectively. Here the long axis is just along the laser polarization, which indicates that the betatron oscillations of electrons are asymmetric in the transverse directions. The background electron plasma densities for nitrogen and helium are 3.0$\times$10$^{18}$ cm$^{-3}$ and 4.6$\times$10$^{18}$cm$^{-3}$, at which the betatron signal are the strongest from each gas. At a plasma density of 3.0$\times$10$^{18}$cm$^{-3}$, no electron signal from helium gas was observed at experiment and simulation. This suggests that the electron beams from nitrogen gas are associated with the K-shell ionization injection process. The electron signal with the nitrogen gas in Fig. 2(i) shows almost continuous spectrum with a spectral peak around 320 MeV and a tail in the low energy. The electron spectrum of helium plasma was also almost continuous with a similar energy peak, but the maximum energy is extended to 500MeV. As exhibited in Fig. 2(k, l), the laser plasma channel lengths with both gas species are approximately the same (~9mm). That means, with similar channel length and lower maximum electron energy, the betatron x-ray yield and photon energy generated from nitrogen were much higher comparing to Helium case. This comparison of results with two gases proves to be important to understand the mechanisms responsible for the enhanced betatron x-ray yield and photon energy from the nitrogen gas, as discussed later.

The spectra of generated betatron x-rays are shown in Figure 3. The average photon flux per keV can be estimated for different photon energy ranges, which are separated by the cut-off energy on different IPs. The average photon numbers for either nitrogen or helium gas are plotted explicitly. For comparison, the detected x-ray spectra in previous work by other groups (Kneip et al.\cite{eric12} and Cipiccia et al.\cite{eric13}) are also shown. One can see that the x-rays generated from helium in our experiment have a similar spectrum as that by Kneip et al in the same laser intensity. However, the x-ray source we obtained from the nitrogen gas has clearly higher photon yields and extends to much higher photon energy. Although, with much lower electron energy in the present case, the photon energy is not as high as the one generated by Cipiccia et al, the nitrogen x-ray source in our case has a much higher photon yield peaked at 30-40keV, which lies in the hard x-ray region. The relative signal strength of the betatron x-ray on IPs 1-4 is illustrated in the inset of Fig. 3, representing the relative experimental x-ray signal strength with nitrogen and helium gases, respectively. The nitrogen betatron x-ray signal is fitted by using a least square fitting method assuming a synchrotron-like spectrum\cite{eric28}. A best fit is found with the critical energy of $E_c=75keV$. With the similar electron energy, the nitrogen betatron x-ray critical energy is more than three times of previous work with He gas\cite{eric12,eric14}. Our spectrum shows an energy peak at 37keV, where the photon flux is about 6.5$\times$10$^6$ photons/keV. The total x-ray yield on spectrum reaches to 8$\times
$10$^{8}$/shot. By integration of the x-ray signal on the last IP, the photon number between 110keV~1MeV is estimated to be $10^8$, which is also in good agreement with the fitted spectrum.

\begin{figure*}[!htb]
\includegraphics[width=0.9\linewidth]{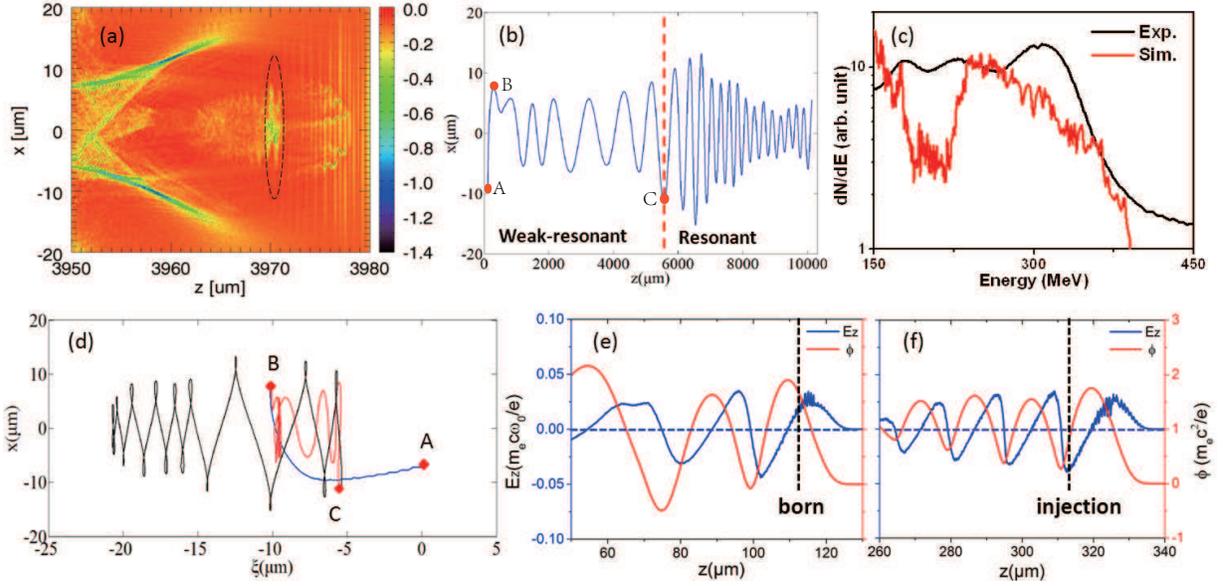}
\caption{\label{4} PIC simulation results. (a) Electron density distribution, where the blue dashed elliptical circle marks the part of electrons performing resonant betatron oscillations; (b) Trajectory of a test electron located within the blue dashed circle; (c) Electron energy spectra from experiment (black) and simulation (red); (d) Trajectory of the same test electron in reference moving frame with $v=v_g$, where $v_g$ is the laser group velocity in the plasma. A, B and C denote the electron born position (released from a nitrogen ion), injection position, and resonant oscillation transition point. The injection, weakly resonant, and strong resonant oscillation processes are shown in blue, red, and black lines, respectively; (e) and (f) show snapshots of the longitudinal wakefield potential (red line) and strength (blue line) at time when the test electron is born and injected, where the black dashed lines in the two plots mark the electron emerging and injection positions, respectively. }
\end{figure*}

To understand the mechanisms for the enhanced betatron radiation in nitrogen gas, two-dimensional (2D) particle-in-cell (PIC) simulations using the KLAPS code\cite{eric29} have been carried out. In the code, field ionization was accounted for with tunnel-ionization model\cite{eric30}. We have used a simulation box of 80$\times$80 $\mu$m$^2$ in size and adopted longitudinal and transverse resolutions of $\Delta$z=0.025$\mu$m and $\Delta$x=0.2$\mu$m, respectively. The neutral nitrogen gas has 100 $\mu$m up-ramp at the front and 100 $\mu$m down-ramp at the rear side, and a 1cm long plateau with the gas density 6$\times$10$^{17}$cm$^{-3}$. A p-polarized 800nm laser pulse with $a_0$=2.2 is focused to a radius of $w_0$=21 $\mu$m at 100$\mu$m after the front edge of the nitrogen gas. The pulse has a Gaussian transverse profile and sine-square longitudinal shape with pulse duration of FWHM=45fs. In the simulation, large amplitudes of betatron oscillation of the trapped electrons are obviously observed as shown in the Fig. 4(a) after 4mm propagation distance of the laser pulse in the nitrogen gas. The trapped electron beam is spatially distributed in the region of the driving laser and it is found to be modulated periodically with a period equal to the local laser wavelength 0.8$\mu$m. The black dashed elliptical circle in Fig. 4(a) highlights this part of electrons, which has the bunch length~10fs. It performs betatron resonant oscillation later. A typical trajectory of a test electron located within this part of trapped bunches in the laboratory frame is shown in Fig. 4(b). The electron firstly experiences small amplitude betatron oscillation due to the original off-axis injection and the restoring force of the wakefield. Because the laser pulse partially fills the bubble, the oscillation amplitude is gradually enlarged during the electron acceleration as a result of interacting weakly with the laser pulse. This process is named as the weak-resonant case. Then the electron is accelerated forward towards the driving laser pulse, and the transverse oscillation amplitude $r_0$ is suddenly enhanced to over 10$\mu$m when it up to the phase rotation point. It's because that the laser frequency is in resonance with a harmonic of the betatron frequency at the phase rotation point\cite{eric13}. In the meanwhile, the oscillation period length is decreased to 400$\mu$m, which corresponds to the frequency of $\omega  = \omega _0 (1 - \frac{{v_z }}{{v_{ph} }})$, with $\omega_{0}$ the laser frequency, $v_{z}$ and $v_{ph}$ the longitudinal velocity of the electron and the phase velocity of the laser pulse, respectively. The high oscillation frequency and large oscillation amplitude result in the enhancement of betatron photon energy and photon yield. It is also proved by elliptical shape of the radiation distribution found in experiment, which suggests the overlap of laser electromagnetic field and the accelerated electron beam. The measured electron spectrum from experiment is also in agreement with the simulation result shown in Fig. 4(c). With this continuous electron energy spectrum, we assume an average electron energy of 200MeV, corresponding to $\gamma$ =400. The resultant betatron x-ray critical energy was estimated to be $
E_c  = 3\hbar a_\beta  \overline \gamma  ^2 \omega _\beta   = 3\hbar c\overline \gamma  ^3 r_0 (2\pi /\overline {\lambda _\beta  } )^2  = 88keV
$
  which was close to the experimental value, assuming a resonant betatron amplitude   $r_0\sim$10$\mu$m, and average betatron oscillation wavelength
$
\overline {\lambda _\beta  }
$
 $\sim$ 400$\mu$m from the simulation.

In order to set insight into injection and acceleration scenario of ionization injection, the trajectory of the same electron as the one in Fig. 4(b) is plotted in a reference frame of the light pulse, as shown in Figure 4(d) where ¦Î denotes the relative distance of electron from the laser peak amplitude. The electron born at point A near the peak of the laser field (¦Î$\sim$0) firstly travels backward relative to the laser pulse up to the point B (via blue line) where it injected. The nonlinear plasma wavelength is 25$\mu$m in the present simulation, while the electron injection point B is just 20$\mu$m from the front of the bubble and 10$\mu$m from the laser peak. This means the electron has been injected well before it moves to the tail of the bubble and very close to the laser pulse. Thus, the injected electron could quickly caught the laser pulse (marked by the red line). At point C, the electron has reached the dephasing phase while the laser field is still strong. Afterwards, the electron moves backward oscillating with large amplitudes and the trajectory is the typical resonant oscillation style forced by the laser electromagnetic field\cite{eric31} (via the black line).
The ionization injection is a kind of fast injection, which enables the betatron oscillations to change from non-resonant to resonant quickly. For an electron born near the peak of the laser field in the plasma wake, the trapping condition could be written as
$
\Delta \varphi  = \varphi _f  - \varphi _i  \le  - 1
$
\cite{eric20,eric32}, where
$
\varphi  = e\Phi /mc^2
$
 is the normalized electrostatic potential of the plasma wave, and the subscripts i and f denote the electron initial ionization and final injection point in the wake, respectively. In the case of a well matched bubble, the maximum wake potential is
$
\varphi _{\max }  \approx (k_p r_b )^2 /4 \approx a_0
$
 , where $k_p$  is the plasma wave number and $r_b$  is the bubble radius. Thus if the electron is born at the point where wake potential is at the maximum, it can be injected when
$
\varphi _f  \le a_0  - 1
$
 . In our case $a_{0}$=2.2 , the electron has a high possibility to be injected with
$
\varphi _f  \le 1.2
$
 . That suggests a fast injection process because the electron can be injected at the middle of bubble and do not need to slip to the bubble tail where the wakefield potential approaches zero. This is confirmed by our simulation. As shown in Figs. 4(e) and 4(f), the wakefield potentials of the ionization and injection point were  $\phi_{i}=1.8$ and $\phi_{f}=0.8$ , these values satisfy the injection condition well. The electron is born at 113$\mu$m and injected at 314$\mu$m, corresponding to a fast injection process of merely 0.66ps.

To realize the betatron resonant condition, the electrons need to approach the phase space rotation point in the laser field\cite{eric13}. For the self-injection case, the electron injection is severely influenced by the nonlinear laser self-focusing process in plasma\cite{eric27,eric33}. Thus the electron injection is usually found at the tail of the bubble where the wakefield potential is minimum. Thus it will take longer time to catch the laser pulse, at which the laser pulse may have decayed drastically because of pump depletion and pulse diffraction. For example, in work\cite{eric13}, a 3cm long capillary was used to obtain the betatron resonant condition and this condition was easily mismatched when the plasma density was slightly changed. On the contrary, with the ionization injection scheme, electrons are injected quickly and close to the laser pulse, so they can reach the phase space rotation point at an earlier time and perform resonant betatron oscillation well before the laser intensity drops down. Thus the fast ionization injection process is beneficial for the occurrence of resonant betatron oscillations. The higher x-ray photon energy and photon yield with nitrogen gas over helium gas in our experiment and previous works\cite{eric12,eric14} support this conclusion. Due to the fast injection process in nitrogen, dephasing takes place earlier and results in lower maximum electron energy compared with helium, which is in good agreeemet with the experimental results in Figs. 2(i) and 2(j).

In conclusion, we demonstrate the generation of intense betatron hard x-ray radiation via ionization injection for the first time by use of 100TW laser pulse incident into a pure nitrogen gas jet. The betatron x-ray has a single pulse photon yield of $8\times10^8$ and has $10^8$ photons over 110keV. The photon energy (critical energy of 75 keV) is much higher and the photon flux is 10 times compared to the case of helium gas (i.e., without the ionization injection) under the same laser parameters. Considering the x-ray duration of 10$fs$, the estimated peak brilliance of the x-ray source is $~10^{23} photons/(mm^2 mrad^2 0.1\%bandwith)$, which is comparable to the 3rd generation synchrotron light source but lies in an unprecedented energetic x-ray region. It is demonstrated that the ionization injection scheme stimulated by nitrogen gas shortens the time for the injected electrons to catch up with the laser pulse, which increases the probability of resonant betatron oscillations. This experiment proved that ionization injection could be a convenient tool to enhance the photon number and energy of betatron x-ray beam from laser-wakefield acceleration. By utilizing PW class laser pulses, stable gamma-ray beams with photon number over $10^{10}$/pulse could be generated from a single stage nitrogen gas jet in the near future.


%
%

%

\begin{acknowledgments}
We gratefully acknowledge to F. Liu and X. L. Ge for the helps on operating the laser facility. This work was supported by the National Basic Research Program of China (2013CBA01500), National Key Scientific Instrument and Equipment Development Project (2012YQ120047), National Natural Science Foundation of China (11334013, 11421064,11374210), the CAS key program (KGZD-EW-T05), and the MOST International Collaboration (2014DFG02330).
\end{acknowledgments}


\end{document}